# Residual Entropy of Ice: A Study Based on Transfer Matrices

De-Zhang Li[1], Yu-Jie Cen[2], Xin Wang[1,3*] and Xiao-Bao Yang[4*]

[1] Quantum Science Center of Guangdong-Hong Kong-Macao Greater Bay Area, Shenzhen 518045, China.

[2] Institute of Materials Chemistry, Vienna University of Technology, Vienna A-1060, Austria.

[3] Department of Physics, City University of Hong Kong, Hong Kong SAR, China.

[4] Department of Physics, South China University of Technology, Guangzhou 510640, China.

[*] Corresponding authors. Correspondence and requests for materials should be addressed to X. W. (email: x.wang@cityu.edu.hk) and X.-B. Y. (email: scxbyang@scut.edu.cn)

## Abstract

The residual entropy of ice systems has long been a significant and intriguing issue in condensed matter physics and statistical mechanics. Exact solutions for the residual entropy of realistic three-dimensional ice systems remain unknown. This study focuses on two typical realistic ice systems: hexagonal ice (ice Ih) and cubic ice (ice Ic). We present a transfer matrix description of the number of ice-ruled configurations for these systems. A transfer matrix $\mathbf{M}$ is constructed for ice Ic, where each element represents the number of ice-ruled configurations of a hexagonal monolayer under certain conditions. The product of $\mathbf{M}$ and its transpose corresponds to a bilayer unit in the ice Ih lattice, thus forming an exact transfer matrix for ice Ih. Utilizing this, we show that the residual entropy of ice Ih is not less than that of ice Ic in the thermodynamic limit, first proved by Onsager in the 1960s. Additionally, we introduce an alternative transfer matrix $\mathbf{M}'$ for ice Ih based on a monolayer periodic unit. Various interesting properties of $\mathbf{M}$, $\mathbf{M}\mathbf{M}^T$ and $\mathbf{M}'$ are analysed, including the summation of all elements, the element in the first row and first column, and the trace. Each property corresponds to the residual entropy of a certain two-dimensional ice model. This work provides an effective description, based on transfer matrices, for the residual entropies of various two-dimensional ice models.

## I. Introduction

Research on ice has garnered significant attention due to its importance in natural sciences [1]. One of its key features is residual entropy, a well-known unresolved problem in condensed matter physics and statistical mechanics. In the 1930s, it was discovered that ice retains nonzero entropy at low temperatures [2, 3]. For this residual entropy, the ice rules [4, 5] offer an explanation. In ice lattices, each oxygen has four nearest neighbors [6], with hydrogens positioned according to the ice rules [7]: (i) there is only one hydrogen between every pair of nearest-neighbor oxygens to form a hydrogen bond; (ii) there are two hydrogens adjacent to each oxygen to constitute an $H_2O$ molecule. Configurations that satisfy these rules are hereafter referred to as "ice-ruled". The disordered arrangements of hydrogens produce numerous configurations, determining the residual entropy

$$S = \frac{1}{N_{\mathrm{H_2O}}} \ln W = \ln w \ , \tag{1}$$

where the Boltzmann constant $k_B$ is fixed as 1, $W$ is the number of ice-ruled configurations, the number of $H_2O$ molecules is denoted by $N_{\mathrm{H_2O}}$, and $w = W^{1/N_{\mathrm{H_2O}}}$. For a hydrogen bond, the direction is defined as that from the donor to the acceptor. Evidently, every ice-ruled hydrogen bond configuration should be two-in/two-out respective to each oxygen site. Thus, $W$ can be considered the number of ways to arrange the ice-ruled hydrogen bonds.

Research on the residual entropy problem of ice models has a long history. Some early studies are as follows: In 1935, Pauling made a Bethe approximation resulting in $w = \frac{3}{2}$ [5], which was later improved by Takahasi [8]. Onsager and Dupuis proved that for four-coordinated ice systems, Pauling's result is a lower bound [9]. In 1964, DiMarzio and Stillinger proposed a matrix method for square ice and three-dimensional ice [10]. Later, in 1966, Nagle presented an advanced estimate using a series expansion method for square ice, hexagonal ice (ice Ih, ordinary ice), and cubic ice (ice Ic) [11, 12]. In 1967, Lieb obtained a well-known exact solution for square ice [13, 14]

$$w = \left(\frac{4}{3}\right)^{\frac{3}{2}} \tag{2}$$

using the transfer matrix approach.

Originating in the study of water ice, residual entropy has become a concept extending beyond hydrogen disorder. In the context of spin ice, the ice rules also apply [15, 16]. More generally, residual entropy arises from extensive ground state degeneracy, with ice-ruled configurations in ice models regarded as the ground states. This makes it a compelling subject in the statistical physics of lattice systems [17]. Examples include the exact results for classical Ising models with geometrical frustration on the triangular [18, 19], Kagomé [20] and checkerboard lattices [21]. There is a close connection between ice models and Ising models, with discussions on the equivalence of ice Ic with the pyrochlore Ising model [22, 23], and square ice with the two-dimensional Ising model [21, 24]. For these statistical models, including the dimer and vertex models, most exactly soluble cases are in one or two dimensions [25]. Several exact solutions exist for the residual entropy of two-dimensional ice models [13, 26-29], while the three-dimensional case remains unsolved. Various research works using theoretical approaches [30-42] and numerical simulations [43-55] have been reported to date.

This study is inspired by Nagle's Ph.D. thesis [56], which proposed an interesting finding: the residual entropy of ice Ih is not less than that of ice Ic. According to the thesis, the proof was provided by Onsager. In addressing the case of three-dimensional ice, a transfer matrix for the number of ice-ruled hydrogen bond configurations in a hexagonal layer was introduced. Onsager discovered the connection between the transfer matrices of ice Ih and ice Ic, thereby demonstrating the relationship in their residual entropies. As is well-known, the transfer matrix method is very useful in the statistical mechanics of lattice models [13, 57-63]. Given that ice Ih and ice Ic are two typical three-dimensional ice systems, we aim to employ the transfer matrix approach based on layer structures in these systems. The main goals of this study are: (i) to generalize the analysis in Nagle's thesis and present a transfer matrix description for the residual entropy of ice Ih and ice Ic; and (ii) to provide an effective representation for various two-dimensional ice models.

The structure of the paper is as follows. In Sec. II, we extend the transfer matrix method used in Nagle's thesis for square ice to construct a monolayer-based transfer matrix $\mathbf{M}$ for ice Ic. We confirm that $\mathbf{M}\mathbf{M}^T$, which corresponds to a bilayer, is a transfer matrix for ice Ih. Subsequently, we demonstrate the relationship $w_{\text{Ih}} \geq w_{\text{Ic}}$. In Sec. III, we introduce an alternative transfer matrix $\mathbf{M}'$ for ice Ih, also based on a monolayer like $\mathbf{M}$. For $\mathbf{M}$, $\mathbf{M}\mathbf{M}^T$ and $\mathbf{M}'$, we illustrate the summation of all elements, the element in the first row and first column, and the trace. Each of these properties is equivalent to the residual entropy of a certain two-dimensional ice model. Sec. IV includes conclusion and discussions.

## II. Transfer Matrix Based on a Monolayer in Ice Ic

In ice Ic, the oxygen sites form a diamond cubic crystal structure. The oxygen lattice can be seen as a stacking of hexagonal layers in an ABCABC pattern, as shown in Fig. 1(a). Each hexagonal monolayer consists of armchair six-membered rings, and the stacking structure is achieved through the vertical bonds between layers. It is evident that each hexagonal monolayer can be treated as a two-dimensional periodic unit of ice Ic by setting the appropriate periodic boundary conditions. Therefore, a transfer matrix $\mathbf{M}$ can be constructed based on the hexagonal monolayer to represent the number of ice-ruled hydrogen bond configurations in ice Ic. Similarly, there is a bilayer periodic unit for ice Ih. We will show that the transfer matrix based on the bilayer of ice Ih is simply $\mathbf{M}\mathbf{M}^T$. Some interesting properties of these two transfer matrices, as well as the solutions of the corresponding two-dimensional ice models, are studied both analytically and numerically.

### A. Construction of M

Consider a hexagonal monolayer with *m* sites. Both the number of lower vertical bonds and that of upper vertical bonds are $\frac{m}{2}$. We label the lower vertical bonds as $1,\ 2,\ 3,\ \ldots\ \frac{m}{2}$ and the upper vertical bonds as $1',\ 2',\ 3',\ \ldots\ \frac{m}{2}'$, as shown in Fig. 1(b) (see also Fig. 11 of Ref. [56]). Let *A* and *B* denote the configurations of the lower vertical bonds and upper vertical bonds, respectively. The element $\mathbf{M}_{A,B}$ of the transfer matrix $\mathbf{M}$ is then defined as the number of ice-ruled hydrogen bond configurations in the layer, given that *A* and *B* are the configurations of the lower and upper vertical bonds, respectively. In this notation *A* refers to the row and *B* refers to the column. Each element of $\mathbf{M}$ is a non-negative integer. The index of the rows of $\mathbf{M}$, i.e., the index for all $2^{m/2}$ possibilities of *A*, can be determined as follows:

$$
\begin{array}{ll}
 & 1,\ \ 2,\ \cdots\ \frac{m}{2}-1,\ \ \frac{m}{2} \\
\text{1st}: & \uparrow,\ \uparrow,\ \cdots\ \uparrow,\ \uparrow \\
\text{2nd}: & \downarrow,\ \uparrow,\ \cdots\ \uparrow,\ \uparrow \\
\text{3rd}: & \uparrow,\ \downarrow,\ \cdots\ \uparrow,\ \uparrow \\
 & \qquad\vdots \\
2^{m/2}-1\ \text{th}: & \uparrow,\ \downarrow,\ \cdots\ \downarrow,\ \downarrow \\
2^{m/2}\,\text{th}: & \downarrow,\ \downarrow,\ \cdots\ \downarrow,\ \downarrow
\end{array}
\tag{3}
$$

Here we use the upward arrow $\uparrow$ and downward arrow $\downarrow$ to represent the configuration of each lower vertical bond. The index of the columns is determined similarly, by changing the vertical bonds from 1, 2, ... $\frac{m}{2}-1$, $\frac{m}{2}$ to $1'$, $2'$, ... $\left(\frac{m}{2}-1\right)'$, $\frac{m}{2}'$. This accomplishes the construction of the $2^{m/2}\times 2^{m/2}$ matrix $\mathbf{M}$. We note that the lower and upper vertical bonds are labelled here in the same way as in Nagle's thesis [56], but the indices of the rows and the columns of $\mathbf{M}$ are not defined there.

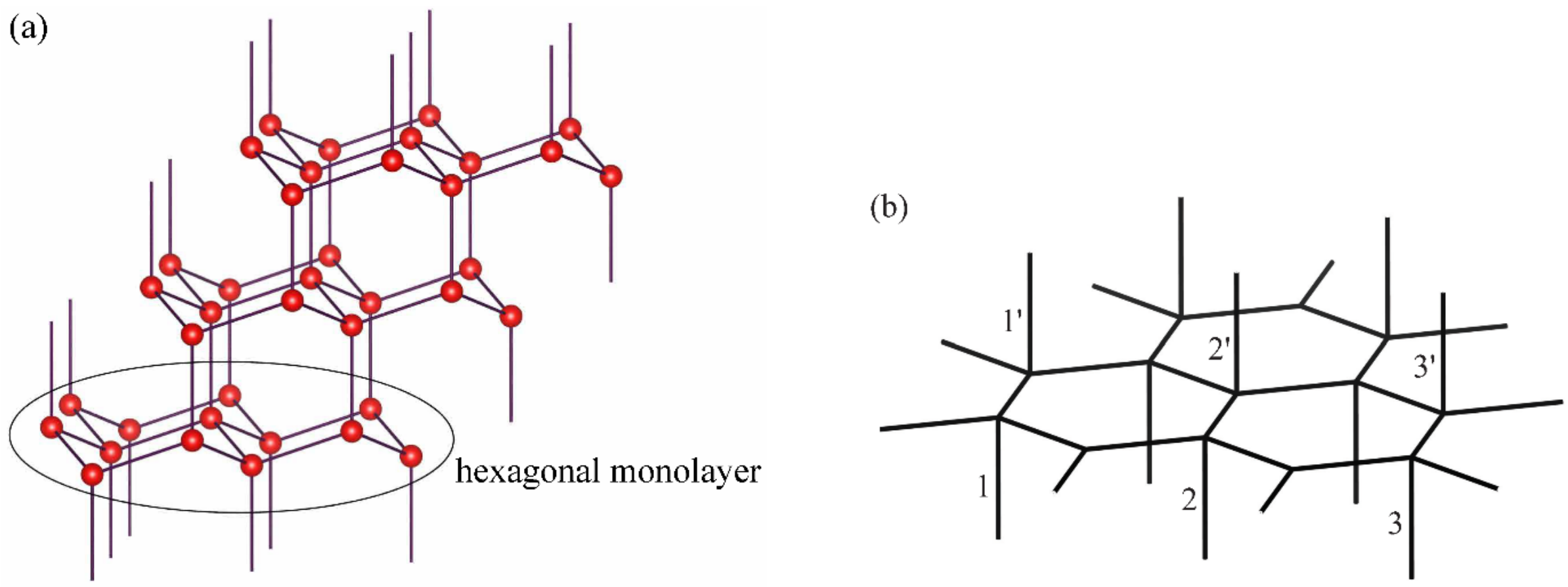

**Fig. 1**. (a) The schematic diagram of the oxygen lattice of ice Ic. Oxygen sites are drawn in red. The hexagonal monolayer as a periodic unit is marked. (b) The labels of the lower and upper vertical bonds of the hexagonal monolayer.

Now one can see that the total number of ice-ruled configurations of $n$ layers in ice Ic is

$$W_{nm}=\mathrm{Tr}\left(\mathbf{M}^{n}\right) \tag{4}$$

under appropriate periodic boundary conditions. The residual entropy in the thermodynamic limit is determined by the largest eigenvalue of $\mathbf{M}$

$$S_{\mathrm{Ic}}=\lim_{n\to\infty,m\to\infty}\frac{1}{nm}\ln W_{nm}=\lim_{n\to\infty,m\to\infty}\frac{1}{nm}\ln\left[\mathrm{Tr}\left(\mathbf{M}^{n}\right)\right]=\lim_{m\to\infty}\frac{1}{m}\ln\left(\lambda_{\max,\mathbf{M}}\right) \tag{5}$$

and $w_{\mathrm{Ic}}$ is simply

$$w_{\mathrm{Ic}}=\lim_{m\to\infty}\lambda_{\max,\mathbf{M}}^{1/m}\ . \tag{6}$$

Although we are not able to exactly solve for the largest eigenvalue $\lambda_{\max,\mathbf{M}}$, it is interesting to compare this value with that of the transfer matrix for ice Ih.

### B. $\mathbf{MM}^{T}$ for Ice Ih

The crystal lattice of the oxygen sites in ice Ih differs from that of ice Ic. The stacking structure of hexagonal layers follows an ABAB pattern, as shown in Fig. 2(a). The vertical bonds link each hexagonal monolayer, chosen as the periodic unit of ice Ic in the preceding section, to its mirror images. It is straightforward to verify that a bilayer, in which a hexagonal monolayer lies beneath its mirror image as marked in Fig. 2(a), can be considered a two-dimensional periodic unit of ice Ih. This two-dimensional structure is parallel to the basal plane of the ice Ih crystal. Therefore, it is natural to introduce a transfer matrix $\mathbf{M}\tilde{\mathbf{M}}$ for ice Ih, where $\mathbf{M}$ corresponds to the bottom layer as stated in the preceding section, and $\tilde{\mathbf{M}}$ is associated with the mirror image layer. The labels for the lower and upper vertical bonds of the mirror image layer are straightforward, as shown in Fig. 2(b). With the indices of the rows and columns determined by Eq. (3), the elements of $\tilde{\mathbf{M}}$ are defined in the same way as those for $\mathbf{M}$. The transfer matrix $\mathbf{M}\tilde{\mathbf{M}}$ is then constructed based on the bilayer, which consists of $2m$ oxygen sites. The upper vertical bonds of this bilayer, whose configurations correspond to the columns of $\mathbf{M}\tilde{\mathbf{M}}$, are now labelled as $1''$, $2''$, $3''$, ... $\frac{m}{2}''$. Each upper vertical bond $i''$ lies directly above $i$.

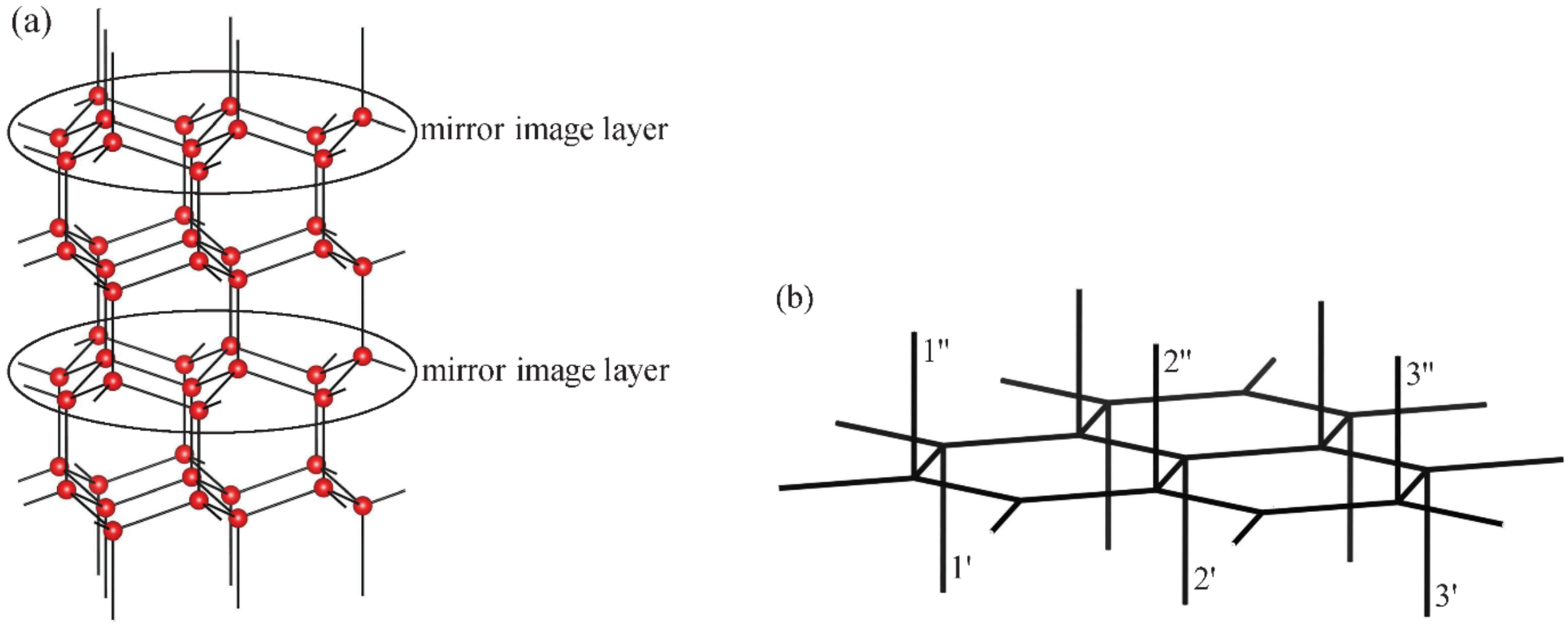

**Fig. 2**. (a). The schematic diagram of the oxygen lattice of ice Ih. The mirror image layer is marked. (b) The labels of the lower and upper vertical bonds of the mirror image layer.

We now demonstrate that $\tilde{\mathbf{M}} = \mathbf{M}^T$. For a configuration $A$, let $-A$ denote its inverse configuration where each bond (arrow) is in the opposite direction of that in $A$. Notice that for every ice-ruled configuration of the mirror image layer, a mirror reflection exists in the bottom layer. That is, there is a one-to-one mapping between the ice-ruled configurations of these two layers. Under the mirror reflection the vertical bonds take the change

$A(\text{lower}), B(\text{upper}) \to -B(\text{lower}), -A(\text{upper})$, thus we have

$$\tilde{\mathbf{M}}_{A,B} = \mathbf{M}_{-B,-A} \ . \tag{7}$$

Since the inverse transformation of the bonds converses the ice-ruled configurations, it is easy to show

$$\mathbf{M}_{-B,-A} = \mathbf{M}_{B,A} \ . \tag{8}$$

Then we know immediately

$$\tilde{\mathbf{M}}_{A,B} = \mathbf{M}_{B,A} \Rightarrow \tilde{\mathbf{M}} = \mathbf{M}^T \ . \tag{9}$$

In this way we have constructed a non-negative symmetric transfer matrix $\mathbf{M}\mathbf{M}^T$ for ice Ih.

As stated in Nagle's thesis [56], Onsager proved that $w_{\text{Ih}} \geq w_{\text{Ic}}$ in the thermodynamic limit making use of the property of a non-negative symmetric matrix. Here we recapitulate this proof. The largest eigenvalue of $\mathbf{M}\mathbf{M}^T$, which is also the square of the 2-norm of $\mathbf{M}^T$, can be expressed by

$$\lambda_{\max,\mathbf{M}\mathbf{M}^T} = \max_{\mathbf{y}} \frac{\mathbf{y}^T \mathbf{M}\mathbf{M}^T \mathbf{y}}{\mathbf{y}^T \mathbf{y}} \ . \tag{10}$$

Denote the leading eigenvector corresponding to the largest eigenvalue of $\mathbf{M}^T$ by $\mathbf{y}_{\max,\mathbf{M}^T}$. Inserting $\mathbf{y}_{\max,\mathbf{M}^T}$ into Eq. (10) gives

$$\lambda_{\max,\mathbf{M}\mathbf{M}^T} \geq \frac{\mathbf{y}^T_{\max,\mathbf{M}^T} \mathbf{M}\mathbf{M}^T \mathbf{y}_{\max,\mathbf{M}^T}}{\mathbf{y}^T_{\max,\mathbf{M}^T} \mathbf{y}_{\max,\mathbf{M}^T}} = \lambda^2_{\max,\mathbf{M}^T} = \lambda^2_{\max,\mathbf{M}} \ . \tag{11}$$

In the last step we use the fact that $\mathbf{M}$ and $\mathbf{M}^T$ have the same eigenvalues. Then the comparison of $w_{\text{Ih}}$ and $w_{\text{Ic}}$ is straightforwardly obtained

$$w_{\text{Ih}} = \lim_{m\to\infty} \lambda^{1/2m}_{\max,\mathbf{M}\mathbf{M}^T} \geq \lim_{m\to\infty} \lambda^{1/m}_{\max,\mathbf{M}} = w_{\text{Ic}} \ . \tag{12}$$

We note that, the fact that $\tilde{\mathbf{M}} = \mathbf{M}^T$ has not been proven in Nagle's thesis [56]. Here we complete the proof.

### C. Properties of $\mathbf{M}$ and $\mathbf{M}\mathbf{M}^T$

For every ice-ruled configuration in a hexagonal monolayer, $\frac{3m}{2}$ hydrogens are in the hexagonal network, and $\frac{m}{2}$ hydrogens are in the vertical bonds. That is, the summation of the number of $\downarrow$s in the lower vertical bonds and that of $\uparrow$s in the upper vertical bonds should be $\frac{m}{2}$. Hence, there should be the same number of $\uparrow$s (also of $\downarrow$s) in the lower vertical bonds and in the upper vertical bonds. This tells us that $\mathbf{M}$ is block diagonal, with the indices of the rows and the columns defined in Eq. (3)

$$\mathbf{M}=\begin{bmatrix} \mathbf{M}_{1,1} & 0 & \cdots & & 0 \\ 0 & [\;] & & & \\ \vdots & & \ddots & & \vdots \\ & & & [\;] & 0 \\ 0 & & \cdots & 0 & \mathbf{M}_{2^{m/2},2^{m/2}} \end{bmatrix}. \tag{13}$$

Obviously $\mathbf{MM}^T$ is also a block diagonal matrix. We now examine three properties of $\mathbf{M}$ and $\mathbf{MM}^T$ in the thermodynamic limit: (i) the summation of all elements, (ii) the element in the first row and first column and (iii) the trace.

**1. M**

The summation of all elements of $\mathbf{M}$ takes into account all $2^m$ configurations of the vertical bonds. This is exactly the residual entropy of the hexagonal monolayer in a zero field, i.e., without any constraint. The exact solution in the thermodynamic limit has been obtained by employing mapping to the antiferromagnetic Kagomé Ising model in our previous work, as shown in Eq. (1) of Ref. [29]. We quote it here

$$\begin{aligned}\lim_{m\to\infty}\frac{1}{m}\ln\left(\sum_{i,j}\mathbf{M}_{i,j}\right)&=\frac{1}{16\pi^2}\int_0^{2\pi}d\theta\int_0^{2\pi}d\phi\;\ln\left\{21-4\left[\cos\theta+\cos\phi+\cos\left(\theta+\phi\right)\right]\right\}\\&=0.752745\;.\end{aligned}\tag{14}$$

As for $\mathbf{M}_{1,1}$, which is equal to $\mathbf{M}_{2^{m/2},2^{m/2}}$, the corresponding model is also easy to find. This element is the number of ice-ruled configurations when all the vertical bonds are $\uparrow$. The model can be thought of the hexagonal monolayer in the presence of a vertical electric field. The residual entropy of this model has been exactly solved, as shown in Eq. (8) of Ref. [29] (see also Sec. IV A of Ref. [28])

$$\begin{aligned}\lim_{m\to\infty}\frac{1}{m}\ln\left(\mathbf{M}_{1,1}\right)&=\frac{1}{16\pi^2}\int_0^{2\pi}d\theta\int_0^{2\pi}d\phi\;\ln\left\{3+2\left[\cos\theta+\cos\phi+\cos\left(\theta+\phi\right)\right]\right\}\\&=0.161533\;.\end{aligned}\tag{15}$$

This model is equivalent to the Kagomé ice [64], of which the ground states can be exactly mapped into the dimer coverings on the honeycomb lattice [26, 27]. Eq. (15), as well as the partition function of the dimer-covering model on the honeycomb lattice [65-67], is one half of the residual entropy of the antiferromagnetic Ising model on the triangular lattice [18, 26,

29, 68].

We now consider the diagonal elements of $\mathbf{M}$. One can see from Eq. (3) that, for each diagonal element the corresponding configurations of the lower and upper vertical bonds are the same, i.e., the bonds of every pair $i$ and $i'$ (see Fig. 1(b)) are in the same direction. It is straightforward to verify that, under this condition $i$ and $i'$ can be converted into one bond connecting the nearest-neighbor oxygen sites $\mathrm{O}_i$ and $\mathrm{O}_{i'}$, on which $i$ and $i'$ are located respectively. Then, the summation of all diagonal elements is exactly the residual entropy of the ice model with double bonds connecting such pairs of nearest-neighbor sites, as shown in Fig. 3(a). In our previous work [29], it was shown that this model can be mapped into the six-vertex model on the square lattice, with the arrow configurations displayed in Fig. 3(b). Each pair of nearest-neighbor sites connected by double bonds is treated as a single site on the square lattice, thus establishing the mapping. It can be easily examined that the vertex weights of the equivalent six-vertex model are

$$\omega_1 = \omega_2 = 1,\ \omega_3 = \omega_4 = 2,\ \omega_5 = \omega_6 = 2\ . \tag{16}$$

The exact result in this case, i.e., the trace of $\mathbf{M}$, is then obtained through the solution of the partition function of this six-vertex model [69-72]. We show the result quoting Eq. (20) of Ref. [29]

$$\begin{aligned}\lim_{m\to\infty}\frac{1}{m}\ln\left[\mathrm{Tr}\left(\mathbf{M}\right)\right] &= \frac{1}{2}\ln 2+\frac{1}{8}\int_{-\infty}^{\infty}\frac{d\alpha}{\cosh\left(\pi\alpha\right)}\ln\left[\frac{\cosh\left(2\mu\alpha\right)-\cos\left(2\mu-\Phi_0\right)}{\cosh\left(2\mu\alpha\right)-\cos\Phi_0}\right] \\ &= 0.473477\ ,\end{aligned} \tag{17}$$

where

$$\mu = \arccos\left(-\frac{1}{4}\right),\ \Phi_0 = \arccos\left(\frac{11}{16}\right)\ . \tag{18}$$

**2. $\mathbf{MM}^T$**

First it is obvious that $\left(\mathbf{MM}^T\right)_{1,1} = \mathbf{M}_{1,1}^2$ from Eq. (13), thus $\lim_{m\to\infty}\frac{1}{2m}\ln\left[\left(\mathbf{MM}^T\right)_{1,1}\right]$ is identical to Eq. (15). The corresponding model is simply the bilayer (see Fig. 2(a)) that all the vertical bonds are $\uparrow$.

The summation of all elements $\sum_{i,j}\left(\mathbf{MM}^T\right)_{i,j}$ is the number of ice-ruled configurations of the bilayer without any constraint on the vertical bonds. Fig. 4(a) shows the structure. The trace $\mathrm{Tr}\left(\mathbf{MM}^T\right)$ counts the number of ice-ruled configurations in which the bonds of every pair $i$ and $i''$ are in the same direction. Once again, we can convert $i$ and $i''$ into one bond connecting $\mathrm{O}_i$ and $\mathrm{O}_{i''}$, thus obtain the model in Fig. 4(b). The model in Fig. 4(a) can be thought of the bilayer with open boundary conditions taken on the lower and upper vertical bonds, while the model in Fig. 4(b) is in the case that periodic boundary conditions are employed.

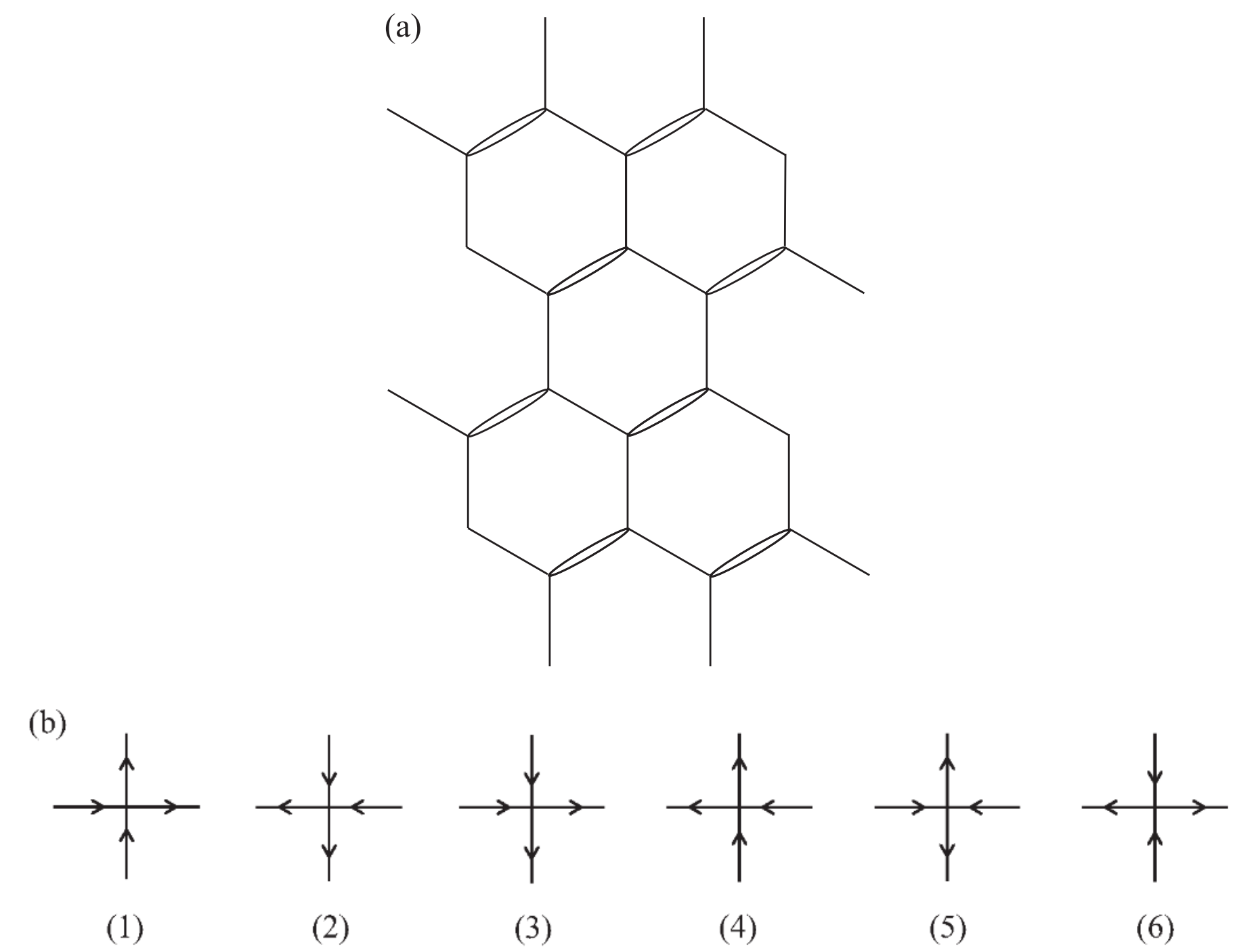

**Fig. 3**. (a) The ice model with double bonds connecting each pair of nearest-neighbor sites $\mathrm{O}_i$ and $\mathrm{O}_{i'}$ in the hexagonal monolayer. (b) The arrow configurations of the six-vertex model.

The value $\sum_{i,j}\left(\mathbf{M}\mathbf{M}^T\right)_{i,j}$ has a mathematically exact lower bound. Taking some simple algebra and employing the Cauchy inequality, we have

$$\begin{aligned}\sum_{i,j}\left(\mathbf{M}\mathbf{M}^T\right)_{i,j} &= \sum_{i,j}\left(\sum_k \mathbf{M}_{i,k}\mathbf{M}^T_{k,j}\right) = \sum_k\sum_{i,j}\mathbf{M}_{i,k}\mathbf{M}_{j,k} \\ &= \sum_k\left(\sum_i \mathbf{M}_{i,k}\right)\left(\sum_j \mathbf{M}_{j,k}\right) = \sum_k\left(\sum_i \mathbf{M}_{i,k}\right)^2 \\ &\geq 2^{-m/2}\left[\sum_k\left(\sum_i \mathbf{M}_{i,k}\right)\right]^2 = 2^{-m/2}\left(\sum_{i,k}\mathbf{M}_{i,k}\right)^2 .\end{aligned} \tag{19}$$

This leads to

$$\lim_{m\to\infty}\frac{1}{2m}\ln\left[\sum_{i,j}\left(\mathbf{M}\mathbf{M}^T\right)_{i,j}\right] \geq \lim_{m\to\infty}\frac{1}{m}\ln\left(\sum_{i,k}\mathbf{M}_{i,k}\right) - \frac{1}{4}\ln 2 = 0.579458 \tag{20}$$

by using Eq. (14). Moreover, we can obtain a mathematically exact lower bound for $w_{\text{Ih}}$, via Eqs. (10), (19) and (14). Denote $(1,\cdots,1)^T$ by $\mathbf{u}$ and the lower bound is simply derived

$$\begin{aligned} w_{\text{Ih}} &= \lim_{m\to\infty} \lambda_{\max,\mathbf{MM}^T}^{1/2m} \geq \lim_{m\to\infty}\left(\frac{\mathbf{u}^T\mathbf{MM}^T\mathbf{u}}{\mathbf{u}^T\mathbf{u}}\right)^{1/2m} \\ &= \lim_{m\to\infty}\left(2^{-m/2}\sum_{i,j}\left(\mathbf{MM}^T\right)_{i,j}\right)^{1/2m} \geq 2^{-1/2}\lim_{m\to\infty}\left(\sum_{i,k}\mathbf{M}_{i,k}\right)^{1/m} \\ &= 2^{-1/2}e^{0.752745} = 1.501060 \ . \end{aligned} \tag{21}$$

As mentioned before, Onsager and Dupuis had pointed out that for four-coordinated ice systems the Bethe approximation $w=\frac{3}{2}$ is actually a lower bound [9]. Now we have obtained an improved lower bound 1.501060 for ice Ih.

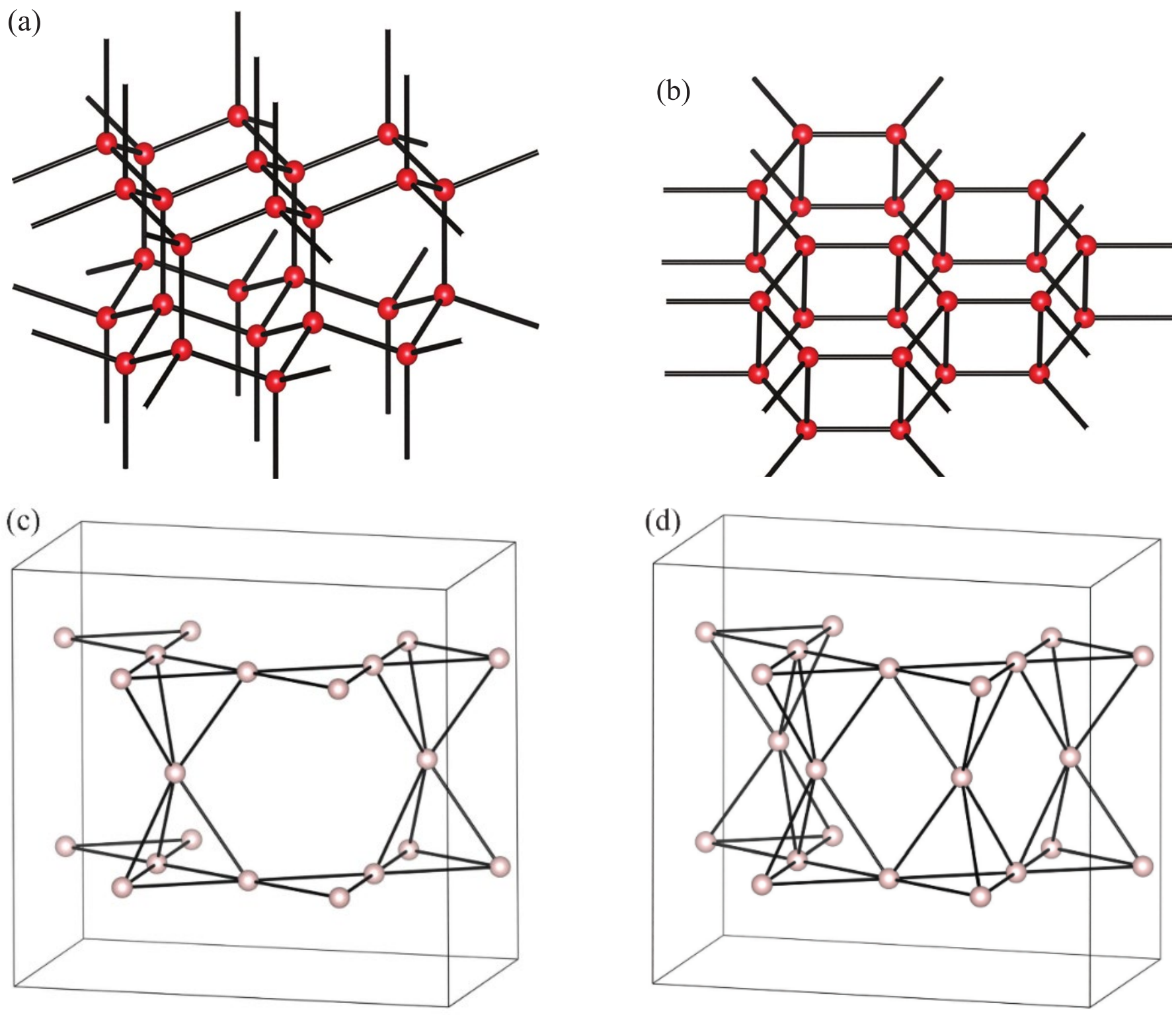

**Fig. 4**. The schematic diagram of the bilayer structures in ice Ih, with (a) open boundary conditions and (b) periodic boundary conditions taken on the lower and upper vertical bonds. (c) and (d) are the orthogonal unit cells of the effective Ising models corresponding to the bilayer structures in (a) and (b), respectively.

For evaluating $\sum_{i,j}\left(\mathbf{MM}^T\right)_{i,j}$ and $\text{Tr}\left(\mathbf{MM}^T\right)$, we perform Wang-Landau Monte Carlo

simulations [73, 74] to estimate the residual entropies of the bilayer structures in Figs. 4(a) and 4(b). In our previous work [55], we have introduced an effective three-dimensional Ising model of which the spin configurations can be exactly mapped into the hydrogen bond configurations of ice Ih. The ground states of the effective Ising model are equivalent with the ice-ruled hydrogen bond configurations, therefore the ground state degeneracy directly determines the residual entropy. Here we employ this approach. Since the bilayer considered here is a two-dimensional unit structure in ice Ih, the construction of the effective Ising models is straightforward (comparing to the three-dimensional case in Ref. [55]). Orthogonal unit cells of the Ising models with nearest-neighbor interactions are used. Figs. 4(c) and 4(d) show the Ising unit cells, which correspond to the bilayer structures in Figs. 4(a) and 4(b), respectively. One can see that each unit cell consists of 8 oxygen sites, and the ground states exactly correspond to the ice-ruled (two-in/two-out) configurations. An improvement of Wang-Landau algorithm [75] is used to simulate the degeneracies of energy states of the effective Ising models with $l^2$ unit cells. Simulations are performed on the systems with sizes $l = 2, 4, \cdots, 16$. For each size, periodic boundary conditions in two dimensions are used and 40 independent samples are generated. $w_N$ is calculated from the ground state degeneracy $w_N = W_N^{1/N}$, with $N = 8l^2$. $w_\infty$ in the thermodynamic limit is obtained from a fit of the data $w_N$ to the expression [44]

$$w_N = w_\infty + a\left(\frac{1}{N}\right)^\theta . \tag{22}$$

For both models, the simulation results of $w_N$ and the fitting values of $w_\infty$ are listed in Table I. Then our estimates are determined directly by $w_\infty$

$$\lim_{m\to\infty}\frac{1}{2m}\ln\left[\sum_{i,j}\left(\mathbf{MM}^T\right)_{i,j}\right] = \ln\left(1.787797 \pm 0.000001\right) = 0.580984 \pm 0.000001 , \tag{23}$$

$$\lim_{m\to\infty}\frac{1}{2m}\ln\left[\mathrm{Tr}\left(\mathbf{MM}^T\right)\right] = \ln\left(1.531063 \pm 0.000010\right) = 0.425962 \pm 0.000006 . \tag{24}$$

The residual entropy of the bilayer structure in Fig. 4(b), i.e., the bilayer model with periodic boundary conditions taken on the lower and upper vertical bonds, has been studied in Ref. [36] (see Fig. 5 there) by a numerical transfer matrix method. Our estimate Eq. (24) is in good agreement with that result 0.4259 ($w = 1.5310$). Actually, this value 1.531063 provides an approximate upper bound for the residual entropy of ice Ih, as it takes all the eigenvalues of $\mathbf{MM}^T$ into account while $w_{\mathrm{Ih}}$ only depends on the largest eigenvalue.

**Table I. The simulation results of $w_N$ and the fitting value of $w_\infty$ for both models.**

| Size $l$ | Number of sites $N$ | $w_N$ of the bilayer in Fig. 4(a) | $w_N$ of the bilayer in Fig. 4(b) |
|---|---|---|---|
| 2 | 32 | $1.797690 \pm 0.000007$ | $1.594298 \pm 0.000011$ |
| 4 | 128 | $1.788054 \pm 0.000005$ | $1.546328 \pm 0.000016$ |
| 6 | 288 | $1.787813 \pm 0.000003$ | $1.537757 \pm 0.000014$ |
| 8 | 512 | $1.787800 \pm 0.000002$ | $1.534788 \pm 0.000012$ |
| 10 | 800 | $1.787797 \pm 0.000002$ | $1.533431 \pm 0.000014$ |
| 12 | 1152 | $1.787797 \pm 0.000002$ | $1.532675 \pm 0.000014$ |
| 14 | 1568 | $1.787800 \pm 0.000002$ | $1.532199 \pm 0.000014$ |
| 16 | 2048 | $1.787803 \pm 0.000002$ | $1.531948 \pm 0.000016$ |
| $\infty$ | $\infty$ | $1.787797 \pm 0.000001$ | $1.531063 \pm 0.000010$ |

Before we end this section, we clarify that the definition of the transfer matrix for ice Ic is not unique. It is obvious that the labels for the lower and upper vertical bonds of the hexagonal monolayer (see Fig. 1(b)) can be determined in other ways, as long as the "transfer" relation between two neighbouring layers is satisfied. In the Appendix we show an alternative construction of the transfer matrix for ice Ic, by changing the labels for vertical bonds.

## III. Transfer Matrix Based on a Monolayer in Ice Ih

Besides the bilayer introduced in Sec. II B, we propose here an alternative two-dimensional periodic unit for ice Ih. We turn to the prism face of ice Ih crystal, and find a hexagonal monolayer parallel to the prism face. Fig. 5(a) shows the monolayer, which is perpendicular to the bilayer in Sec. II B and consists of boat six-membered rings. One sees that under appropriate periodic boundary conditions, this hexagonal monolayer can also be a periodic unit of ice Ih since the layer-by-layer "transfer" relation is satisfied. Therefore, an alternative transfer matrix $\mathbf{M}'$ to represent the number of ice-ruled configurations of ice Ih can be built up. Also, we examine three properties of $\mathbf{M}'$ and the corresponding two-dimensional ice models as we have done for $\mathbf{M}$ and $\mathbf{M}\mathbf{M}^T$.

### A. Construction of $\mathbf{M}'$

The construction of $\mathbf{M}'$ is very similar to that of $\mathbf{M}$. We still let there be *m* sites in the monolayer, and name the bonds linking two neighbouring layers the "lower" and "upper" bonds. $\frac{m}{2}$ lower bonds and $\frac{m}{2}$ upper bonds are in a different arrangement, comparing to the lower and upper vertical bonds of the monolayer corresponding to $\mathbf{M}$. The labels $1,\ 2,\ 3,\ \ldots\ \frac{m}{2}$ for the lower bonds and $1',\ 2',\ 3',\ \ldots\ \frac{m}{2}'$ for the upper bonds are shown in Fig. 5(b). The configurations of the lower bonds refer to the rows of $\mathbf{M}'$, while those of the upper bonds refer to the columns. The indices of the rows and the columns are still defined in Eq. (3), and each element $\mathbf{M}'_{A,B}$ is the number of ice-ruled configurations in the layer when $A$ and $B$ are the configurations of the lower and upper bonds, respectively.

Comparing Fig. 5(b) with Fig. 1(b) clearly shows that, $\mathbf{M}'$ is formed by rearranging the elements of $\mathbf{M}$. Since $\mathbf{M}'$ is a transfer matrix for ice Ih, $w_{\mathrm{Ih}}$ can also be expressed by the largest eigenvalue of $\mathbf{M}'$

$$w_{\mathrm{Ih}} = \lim_{m\to\infty} \lambda_{\max,\mathbf{M}'}^{1/m} \ . \tag{25}$$

Therefore, the relationship between $w_{\mathrm{Ih}}$ and $w_{\mathrm{Ic}}$ (Eq. (12)) leads to

$$\lim_{m\to\infty} \lambda_{\max,\mathbf{M}'}^{1/m} = \lim_{m\to\infty} \lambda_{\max,\mathbf{M}\mathbf{M}^T}^{1/2m} \geq \lim_{m\to\infty} \lambda_{\max,\mathbf{M}}^{1/m} \ . \tag{26}$$

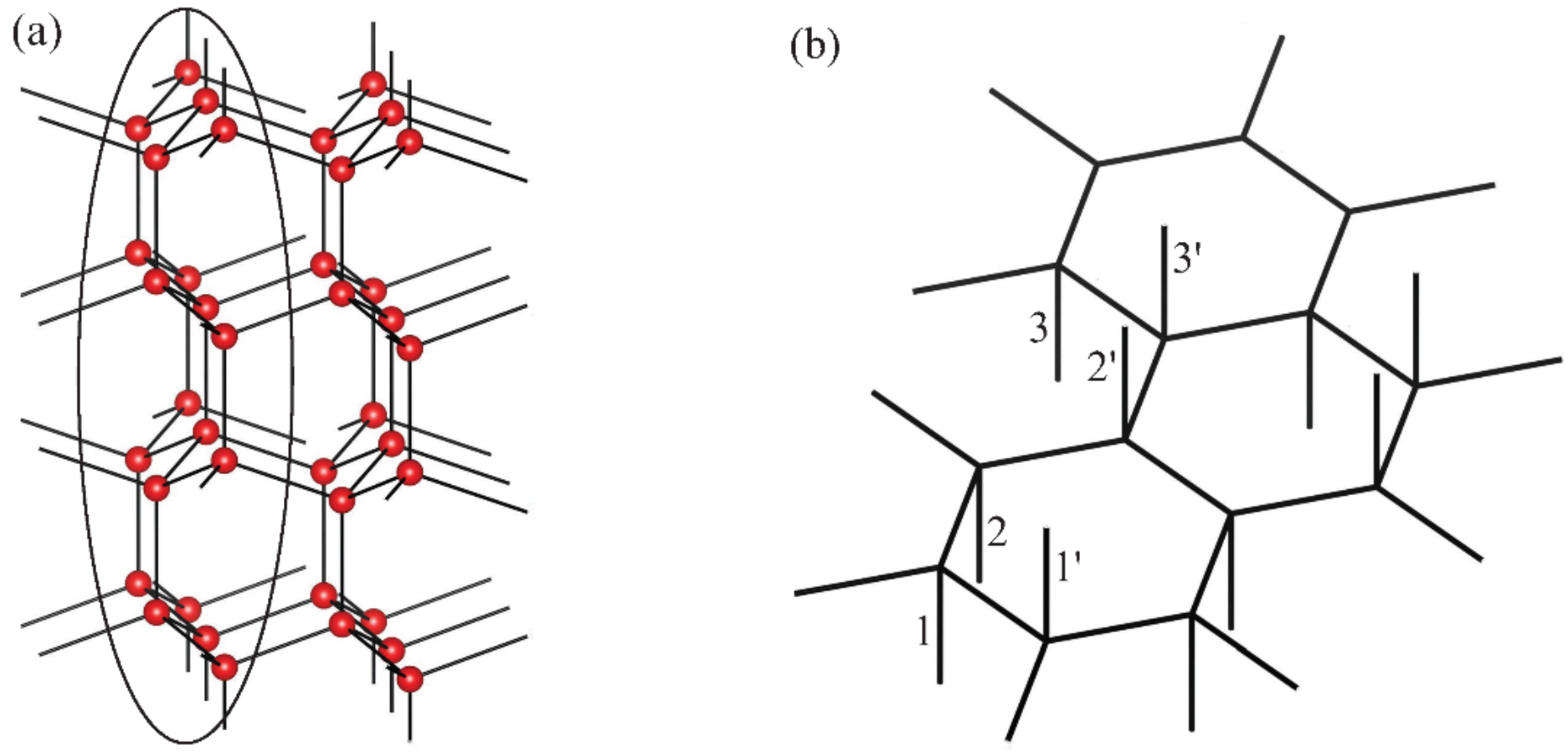

**Fig. 5**. (a) The schematic diagram of the monolayer as a periodic unit of ice Ih. (b) The labels of the lower and upper bonds of the monolayer.

## B. Properties of $\mathbf{M}'$

Using the similar analysis as that for $\mathbf{M}$, we know for every ice-ruled configuration the number of $\uparrow$s (also of $\downarrow$s) in the lower bonds is the same with that in the upper bonds.

Obviously $\mathbf{M}'$ is also a block diagonal matrix taking the form of Eq. (13).

As pointed out, the construction of $\mathbf{M}'$ can be thought of a rearrangement of the elements of $\mathbf{M}$. The summation of all elements of $\mathbf{M}'$ in the thermodynamic limit $\lim_{m\to\infty}\frac{1}{m}\ln\left(\sum_{i,j}\mathbf{M}'_{i,j}\right)$ is thus completely identical to that of $\mathbf{M}$ in Eq. (14). The corresponding model is also consistent with the case there, i.e., the two-dimensional ice model on the hexagonal network.

$\mathbf{M}'_{1,1}$ is the number of ice-ruled configurations under the condition that all the lower and upper bonds are $\uparrow$. Ref. [28] analysed a case that an electric field is applied along [010] direction in ice Ih and the system is reduced to decoupled layers. Each layer in that case is equivalent with our model considered here. An exact mapping from the layer in that case to the dimer model on the square lattice is proposed in Sec. IV B of Ref. [28]. Here we give a reexplanation briefly. One sees from Fig. 5(b) that the configuration of three bonds in the hexagonal network should be one-in/two-out respective to $\mathrm{O}_i$, and two-in/one-out respective to $\mathrm{O}_{i'}$. Denote the pair of nearest-neighbor sites $\mathrm{O}_i$ and $\mathrm{O}_{i+1}$ by $\mathrm{O}_i-\mathrm{O}_{i+1}$ (e.g., $\mathrm{O}_1-\mathrm{O}_2$), or $\mathrm{O}_{i'}$ and $\mathrm{O}_{i+1'}$ by $\mathrm{O}_{i'}-\mathrm{O}_{i+1'}$ (e.g., $\mathrm{O}_{2'}-\mathrm{O}_{3'}$). By regarding such pair $\mathrm{O}_i-\mathrm{O}_{i+1}$ or $\mathrm{O}_{i'}-\mathrm{O}_{i+1'}$ as a single site, a square lattice with two sublattices is obtained. The configuration of four bonds around the site $\mathrm{O}_i-\mathrm{O}_{i+1}$ on the square lattice is one-in/three-out, while that for $\mathrm{O}_{i'}-\mathrm{O}_{i+1'}$ is three-in/one-out. Thus, the pairs $\mathrm{O}_i-\mathrm{O}_{i+1}$ and $\mathrm{O}_{i'}-\mathrm{O}_{i+1'}$ form sublattices A and B, respectively. Each site on A (one-in/three-out) is surrounded by four sites on B (three-in/one-out), and vice versa. Clearly, for every site on A (B) there is exactly one bond pointing inward (outward). Then, by regarding such bonds $\mathrm{B}\to\mathrm{A}$ as dimers, a one-to-one mapping from the bond configurations to the dimer-coverings on the square lattice can be found. Now we see, $\mathbf{M}'_{1,1}$ (and $\mathbf{M}'_{2^{m/2},2^{m/2}}$) is identical to the partition function of the dimer-covering model on the square lattice [67, 76-80]

$$\begin{aligned}\lim_{m\to\infty}\frac{1}{m}\ln\left(\mathbf{M}'_{1,1}\right)&=\frac{1}{2}\times\frac{1}{16\pi^2}\int_0^{2\pi}d\theta\int_0^{2\pi}d\phi\ \ln\left[4+2\cos\theta+2\cos\phi\right]\\&=\frac{G}{2\pi}\\&=0.145780\ .\end{aligned}\tag{27}$$

Here the factor $\frac{1}{2}$ arises from the ratio that two oxygens correspond to one site on the square lattice, and $G$ is Catalan's constant.

The trace of $\mathbf{M}'$ is the number of ice-ruled configurations in which the bonds of every pair $i$ and $i'$ are in the same direction. Again, such bonds of a pair $i$ and $i'$ are converted into one bond connecting the nearest-neighbor sites $\mathrm{O}_i$ and $\mathrm{O}_{i'}$, thereby forming the ice model in Fig. 6(a). $\mathrm{Tr}(\mathbf{M}')$ is exactly the residual entropy of this model. One can easily find that this model is comparable to the one in Fig. 3(a), of which the residual entropy is $\mathrm{Tr}(\mathbf{M})$ (Eq. (17)). Using the same technique with that for the case in Fig. 3(a), we map this model into a six-vertex model on the square lattice by regarding each pair of nearest-neighbor oxygens connected by double bonds as a single site. Then we find that the obtained six-vertex model exhibits a "row-by-row" staggered structure, with two sublattices $L$ and $L'$ as shown in Fig. 6(b). The arrow configurations are still those in Fig. 3(b), and the vertex weights of the sites on two sublattices are

$$\begin{aligned} &\omega_1 = \omega_2 = 1,\ \omega_3 = \omega_4 = 2,\ \omega_5 = \omega_6 = 2 \qquad \text{on } L \\ &\omega_1' = \omega_2' = 2,\ \omega_3' = \omega_4' = 1,\ \omega_5' = \omega_6' = 2 \qquad \text{on } L' . \end{aligned} \tag{28}$$

$\mathrm{Tr}(\mathbf{M}')$ is the partition function of this six-vertex model. The weights of the sites on $L$ are the same with those of the model corresponding to Fig. 3(a) (see Eq. (16)), therefore we denote that model by $(L)$ for convenience. The "row-by-row" staggered model considered here is thus denoted by $(L+L')$. We will illustrate that the partition function of $(L+L')$ is identical to that of $(L)$ in the thermodynamic limit.

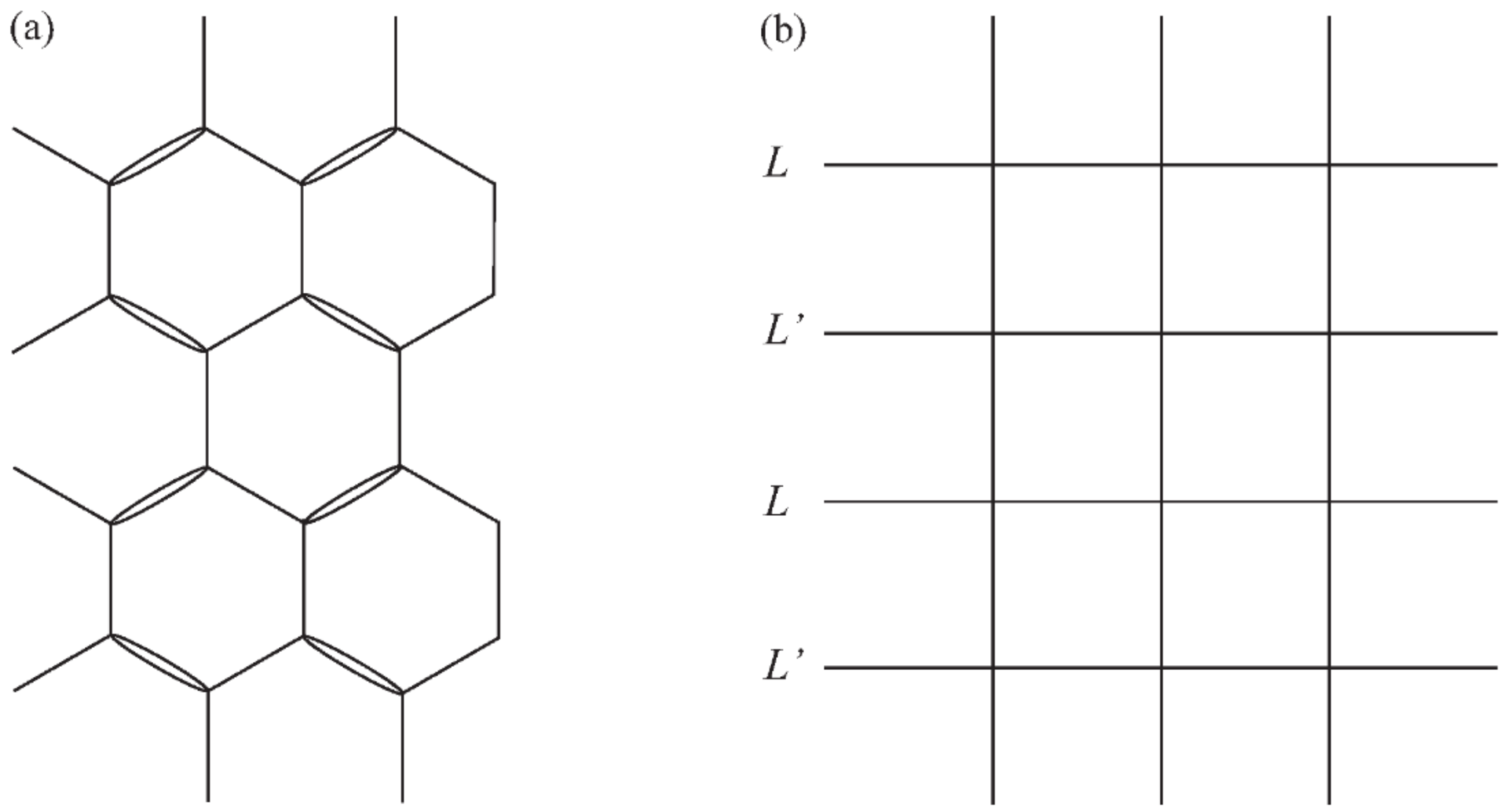

**Fig. 6**. (a) The ice model with double bonds connecting each pair of nearest-neighbor sites $\mathrm{O}_i$ and $\mathrm{O}_{i'}$ in the monolayer. (b) The square lattice of the "row-by-row" staggered six-vertex

model.

Solutions of staggered six-vertex models have been studied both analytically and numerically [81-92]. Here we analyse the row-by-row transfer matrix of $\left(L+L'\right)$ in the case that

$$\begin{aligned} &\omega_1=\omega_2=a,\ \omega_3=\omega_4=b,\ \omega_5=\omega_6=c \qquad \text{on } L \\ &\omega_1'=\omega_2'=b,\ \omega_3'=\omega_4'=a,\ \omega_5'=\omega_6'=c \qquad \text{on } L'\ . \end{aligned} \tag{29}$$

Let there be *p* vertices in a row with periodic boundary conditions and assume *a*, *b* and *c* are positive. Consider a row of vertices in the lattice with *p* upper vertical arrows and *p* lower vertical arrows. Denote the configurations of the upper and lower vertical arrows by *A* and *B*, respectively, and the set of allowed configurations in the row when *A* and *B* are given by $\left\{\sigma\left(A,B\right)\right\}$. Then we can define the transfer matrix $\mathbf{V}$ on $L$ with the elements

$$\mathbf{V}_{A,B}=\sum_{\left\{\sigma\left(A,B\right)\right\}}\prod_{i=1}^{6}\omega_i^{\sigma_i\left(A,B\right)}\ , \tag{30}$$

where $\sigma_i\left(A,B\right)$ is the number of vertices (*i*) in the allowed configuration $\sigma\left(A,B\right)$. $\mathbf{V}'$ on $L'$ is defined in the same way. The transfer matrix for $\left(L\right)$ is thus $\mathbf{V}$, and that for $\left(L+L'\right)$ is $\mathbf{V}\mathbf{V}'$. There are three possibilities for the number of allowed configurations in the row:

<1>. When $A=B$, the horizontal arrows are either all $\rightarrow$ or all $\leftarrow$ in the allowed configuration. The number is 2. In this case $\mathbf{V}_{A,A}$, i.e., the diagonal element of $\mathbf{V}$, is in the form of $\mathbf{V}_{A,A}=\omega_1^x\omega_3^{p-x}+\omega_2^{p-x}\omega_4^x$.

<2>. When $A\neq B$, for certain vertices the upper and lower vertical arrows are in the opposite direction. Denote these two types of vertices by $\substack{\uparrow\\ \downarrow}$ and $\substack{\downarrow\\ \uparrow}$. If $\substack{\uparrow\\ \downarrow}$ and $\substack{\downarrow\\ \uparrow}$ interlace in the row, i.e., $\substack{\uparrow\cdots\downarrow\cdots\uparrow\cdots\downarrow\cdots\\ \downarrow\cdots\uparrow\cdots\downarrow\cdots\uparrow\cdots}$, the horizontal arrows between $\substack{\uparrow\\ \downarrow}$ (left) and $\substack{\downarrow\\ \uparrow}$ (right) are $\leftarrow$, and those between $\substack{\downarrow\\ \uparrow}$ (left) and $\substack{\uparrow\\ \downarrow}$ (right) are $\rightarrow$. In this case the number is 1. $\sigma_5\left(A,B\right)=\sigma_6\left(A,B\right)$ and $\mathbf{V}_{A,B}$ takes a form of $\mathbf{V}_{A,B}=\omega_1^{\sigma_1\left(A,B\right)}\omega_2^{\sigma_2\left(A,B\right)}\omega_3^{\sigma_3\left(A,B\right)}\omega_4^{\sigma_4\left(A,B\right)}\left(\omega_5\omega_6\right)^{\sigma_5\left(A,B\right)}$.

<3>. The number is 0, otherwise.

When we exchange *A* and *B*, these three cases can also be easily examined. Case <1> is trivial.

In case <2>, there is still one allowed configuration. All the horizontal arrows should be inverse, which leads to the transformation of vertices

$$(1)\leftrightarrow(4),\ (2)\leftrightarrow(3),\ (5)\leftrightarrow(6)\ . \tag{31}$$

In case <3>, there is still no allowed configuration. Now we elaborate the element $\mathbf{V}'_{A,B}$. In case <1>, we have

$$\mathbf{V}'_{A,A} = \omega_1'^{x}\omega_3'^{p-x} + \omega_2'^{p-x}\omega_4'^{x} = \omega_4^{x}\omega_2^{p-x} + \omega_3^{p-x}\omega_1^{x} = \mathbf{V}_{A,A}\ . \tag{32}$$

In case <2>, it can be shown that

$$\begin{aligned}\mathbf{V}'_{A,B} &= \omega_1'^{\sigma_1(A,B)}\omega_2'^{\sigma_2(A,B)}\omega_3'^{\sigma_3(A,B)}\omega_4'^{\sigma_4(A,B)}\left(\omega_5'\omega_6'\right)^{\sigma_5(A,B)} \\ &= \omega_4^{\sigma_4(B,A)}\omega_3^{\sigma_3(B,A)}\omega_2^{\sigma_2(B,A)}\omega_1^{\sigma_1(B,A)}\left(\omega_5\omega_6\right)^{\sigma_6(B,A)} \\ &= \mathbf{V}_{B,A}\ .\end{aligned} \tag{33}$$

In case <3>, $\mathbf{V}'_{A,B} = \mathbf{V}_{B,A} = 0$. Therefore, $\mathbf{V}' = \mathbf{V}^T$ and the transfer matrix for $\left(L + L'\right)$ is $\mathbf{V}\mathbf{V}^T$.

It is well known that the eigenvectors of the transfer matrix of a six-vertex model are exactly the same with those of a Heisenberg XXZ model [13, 25, 69, 93, 94], based on the Bathe ansatz. The leading eigenvector of the transfer matrix associated with the largest eigenvalue is identical to the ground state of the XXZ model. The corresponding XXZ model of $\mathbf{V}$ is

$$H = -\sum_{i=1}^{p}\left(S_i^x S_{i+1}^x + S_i^y S_{i+1}^y + \Delta S_i^z S_{i+1}^z\right) \tag{34}$$

with the anisotropy parameter $\Delta$ determined by the weights

$$\Delta = \frac{a^2 + b^2 - c^2}{2ab}\ . \tag{35}$$

One can see clearly this is also the corresponding XXZ model of $\mathbf{V}'$ because $\Delta' = \Delta$ from Eq. (29). Then it turns out that $\mathbf{V}$ and $\mathbf{V}'$ have the same eigenvectors exactly, and also the same leading eigenvector. Moreover, the fact $\mathbf{V}' = \mathbf{V}^T$ tells us that they have the same eigenvalues. Now we can express $\mathbf{V}$ and $\mathbf{V}^T$ as $\mathbf{V} = \mathbf{Q}\boldsymbol{\Lambda}\mathbf{Q}^{-1}$ and $\mathbf{V}^T = \mathbf{Q}\boldsymbol{\Lambda}'\mathbf{Q}^{-1}$, respectively. Here $\boldsymbol{\Lambda}$ and $\boldsymbol{\Lambda}'$ are diagonal matrices consisting of the eigenvalues $\left\{\lambda_{i,\mathbf{V}}\right\}$ and $\mathbf{Q}$ is comprised of the eigenvectors. Therefore, we have

$$\mathbf{V}\mathbf{V}^T = \mathbf{Q}\boldsymbol{\Lambda}\boldsymbol{\Lambda}'\mathbf{Q}^{-1}\ . \tag{36}$$

Each eigenvalue of $\mathbf{V}\mathbf{V}^T$ is in the form of $\lambda_{i,\mathbf{V}}\lambda_{j,\mathbf{V}}$, hence is not larger than $\lambda^2_{\max,\mathbf{V}}$. Since $\mathbf{V}$ and $\mathbf{V}^T$ have the same leading eigenvector, $\lambda^2_{\max,\mathbf{V}}$ is one of the eigenvalues. Then we know that, the largest eigenvalue of $\mathbf{V}\mathbf{V}^T$ must be $\lambda^2_{\max,\mathbf{V}}$, i.e.,

$$\lambda_{\max,\mathbf{V}\mathbf{V}^T} = \lambda^2_{\max,\mathbf{V}} \ . \tag{37}$$

Now we can conclude that, the partition functions of $(L+L')$ and $(L)$ in the thermodynamic limit are identical

$$\lim_{N_{\text{site}}\to\infty} \frac{1}{N_{\text{site}}} \ln Z_{(L+L')} = \lim_{p\to\infty} \frac{1}{2p} \ln\left(\lambda_{\max,\mathbf{V}\mathbf{V}^T}\right) = \lim_{p\to\infty} \frac{1}{p} \ln\left(\lambda_{\max,\mathbf{V}}\right) = \lim_{N_{\text{site}}\to\infty} \frac{1}{N_{\text{site}}} \ln Z_{(L)} \ . \tag{38}$$

That is, the exact solution of the "row-by-row" staggered six-vertex model $(L+L')$ with the weights in Eq. (29) is simply the same with that of $(L)$ (or $(L')$).

As a special case of this finding, we confirm that the exact results of $\mathrm{Tr}(\mathbf{M}')$ and $\mathrm{Tr}(\mathbf{M})$ are equal (see Eq. (17))

$$\lim_{m\to\infty} \frac{1}{m} \ln\left[\mathrm{Tr}(\mathbf{M}')\right] = \lim_{m\to\infty} \frac{1}{m} \ln\left[\mathrm{Tr}(\mathbf{M})\right] = 0.473477 \ . \tag{39}$$

We remark that the model in Fig. 6(a) has been constructed previously in Ref. [35] in a different context, called the digonal-hexagonal ice model (see Fig. 5(b) there). A numerical transfer matrix method was used in Ref. [35] to calculate the residual entropy, and the result of extrapolation to infinite system is 0.473498 ( $w = 1.6056$ ). This estimate is in excellent agreement with the exact solution Eq. (39).

## IV. Conclusion and Discussions

In this paper, we have presented a transfer matrix description for the residual entropy of ice Ih and ice Ic. Each transfer matrix, representing the number of ice-ruled configurations, is based on a layer structure that can be regarded as the periodic unit of the lattice system. Similar to the case of the classical partition function, the residual entropy calculation in the thermodynamic limit can be formulated as the problem of evaluating the largest eigenvalue. A hexagonal monolayer is chosen as the periodic unit of ice Ic, and a transfer matrix $\mathbf{M}$ is constructed based on this monolayer. By realizing a bilayer periodic unit that consists of the hexagonal monolayer and its mirror image, we indicate the corresponding transfer matrix $\mathbf{M}\mathbf{M}^T$ for ice Ih. We further find that a monolayer parallel to the prism face can also be a periodic unit of ice Ih, and thus propose an alternative transfer matrix $\mathbf{M}'$.

Although we are not able to exactly solve for the largest eigenvalues, we illustrate some interesting properties of these transfer matrices and provide solutions for various two-dimensional ice models. We use both analytical and numerical methods, employing mappings from ice models to Ising models, dimer models, and the six-vertex model. In particular, we highlight two achievements of this paper:

The first is the proof that the residual entropy of ice Ih is not less than that of ice Ic in the thermodynamic limit. We extend the method in Nagle's thesis [56] to construct the transfer matrices $\mathbf{M}$ and $\mathbf{MM}^T$ for ice Ic and ice Ih, respectively. The relationship in the largest eigenvalues $\lambda_{\max,\mathbf{MM}^T} \geq \lambda_{\max,\mathbf{M}}^2$ simply leads to $w_{\mathrm{Ih}} \geq w_{\mathrm{Ic}}$, thereby rediscovering Onsager's proof.

The second is the exact result of $\mathrm{Tr}\left(\mathbf{M}'\right)$, which is the residual entropy of the two-dimensional ice model in Fig. 6(a). This model has been previously constructed, and a numerical estimate was obtained [35]. We present the exact solution by mapping this model to a "row-by-row" staggered six-vertex model with the weights given by Eq. (28). More generally, we analyse the "row-by-row" staggered six-vertex model $\left(L+L'\right)$ in the case of Eq. (29), and prove in detail that its partition function is exactly the same as that of $\left(L\right)$ or $\left(L'\right)$ in the thermodynamic limit. The proof is given by applying the properties of the row-by-row transfer matrices: (i) $\mathbf{V}' = \mathbf{V}^T$ and (ii) $\mathbf{V}$ and $\mathbf{V}'$ have exactly the same eigenvectors and the same leading eigenvector. This result motivates us to consider staggered vertex models in more general cases, such as the staggered eight-vertex model [95-97] with the "row-by-row" weights given by

$$\begin{aligned} &\omega_1 = \omega_2 = a,\ \omega_3 = \omega_4 = b,\ \omega_5 = \omega_6 = c,\ \omega_7 = \omega_8 = d \qquad \text{on } L \\ &\omega_1' = \omega_2' = b,\ \omega_3' = \omega_4' = a,\ \omega_5' = \omega_6' = c,\ \omega_7' = \omega_8' = d \qquad \text{on } L' \ . \end{aligned} \tag{40}$$

Finally, we note that solving exactly the bilayer models in Fig. 4, or equivalently the Ising models on a stacked Kagomé bilayer lattice, is challenging. The Wang-Landau algorithm is a useful numerical tool, and we employ it to simulate the ground state degeneracy. In fact, exact solutions of frustrated systems remain difficult to achieve, even for two-dimensional models. Theoretical approximations and numerical computations are necessary. The application of advanced theoretical approaches, such as the tensor network methods [98], to these complicated statistical models is currently in progress.

## Statements and Declarations

### Acknowledgements

This work was supported by Guangdong Provincial Quantum Science Strategic Initiative (Grant No. GDZX2203001, GDZX2303007), Guangdong Basic and Applied Basic Research Foundation (Grant No. 2021A1515010328), National Natural Science Foundation of China (Grant No. 11874312 and No. 12074126), the Research Funding for Outbound Postdoctoral Fellows in Shenzhen (Grant No. SZRCXM2401006), and the Innovation Program for Quantum Science and Technology (Grant No. 2021ZD0302300).

## Appendix: An Alternative Transfer Matrix for Ice Ic

The transfer matrix $\mathbf{M}$ for ice Ic is constructed based on the hexagonal monolayer, with the labels for the lower and upper vertical bonds shown in Fig. 1(b). Here we present an alternative transfer matrix $\mathbf{M}''$ for ice Ic. $\mathbf{M}''$ is built up based on the same monolayer consisting of $m$ sites, but with different labels for vertical bonds as shown in Fig. 7. We can see that the layer-by-layer "transfer" relation is still satisfied when the new labels are used, although the periodic boundary conditions are changed. Therefore, $\mathbf{M}''$ can be a transfer matrix representing the number of ice-ruled configurations. The indices of the row and the column are still defined by Eq. (3).

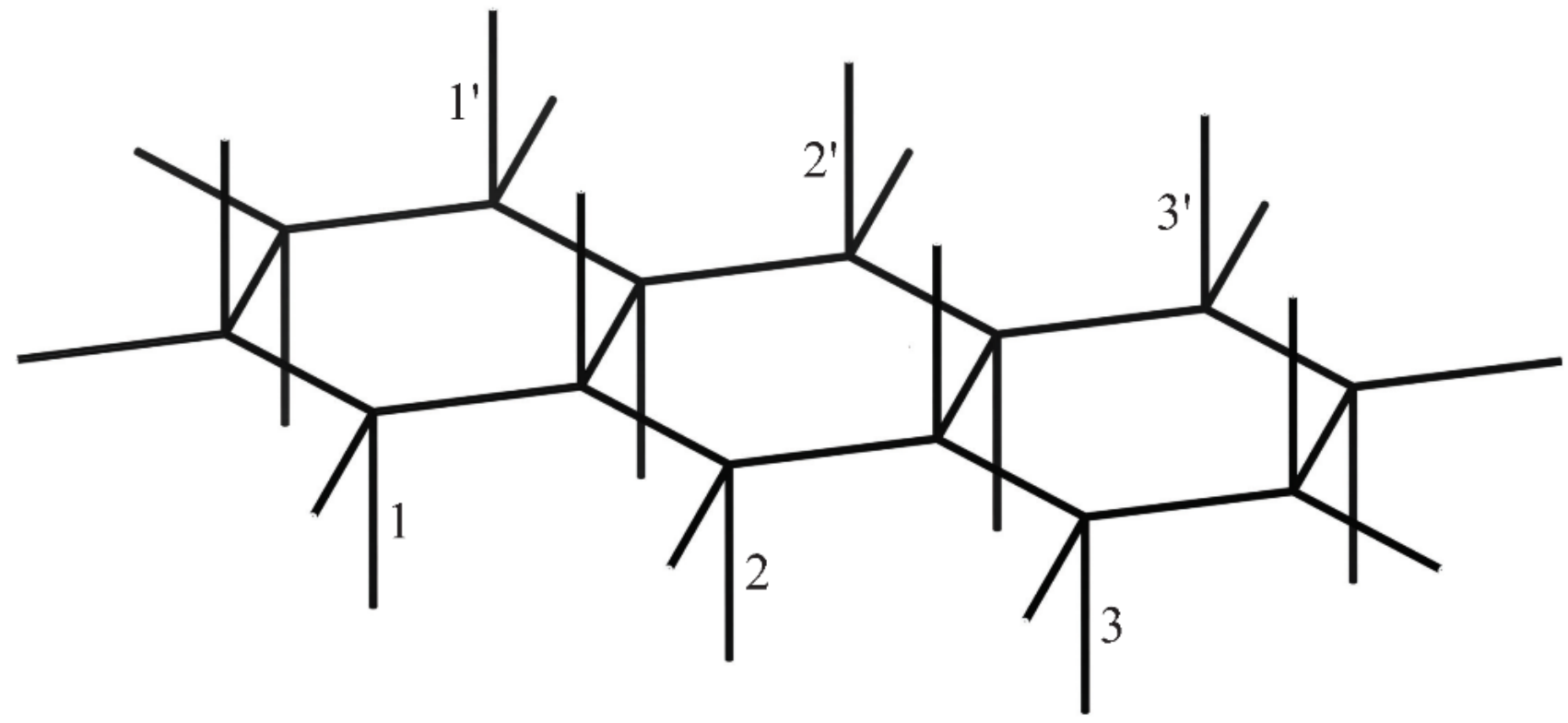

**Fig. 7**. Alternative labels for the lower and upper vertical bonds of the hexagonal monolayer.

By comparing Fig. 7 with Fig. 1(b), we know that $\mathbf{M}''$ can be viewed as a rearrangement of the elements of $\mathbf{M}$, like $\mathbf{M}'$ does. $\mathbf{M}''$ is also a block diagonal matrix in the form of Eq. (13). The summation of all elements of $\mathbf{M}''$ is the same with that of $\mathbf{M}$, shown in Eq. (14). We can also straightforwardly see that $\mathbf{M}''_{1,1} = \mathbf{M}_{1,1}$, as the corresponding models are consistent, i.e., the hexagonal monolayer that all the vertical bonds are $\uparrow$. Thus $\lim_{m\to\infty} \frac{1}{m} \ln\left(\mathbf{M}''_{1,1}\right)$ is equal to Eq. (15). The result of $\mathrm{Tr}\left(\mathbf{M}''\right)$, i.e., the number of ice-ruled configurations in which the bonds of every pair $i$ and $i'$ are in the same direction, is interesting. Again, we can find the corresponding model by converting the bonds of each pair $i$ and $i'$ into a bond connecting $O_i$ and $O_{i'}$. From Fig. 7 we see the resulting model in this case is the square ice, which can be seen as a special case of the six-vertex model where $\omega_1 = \cdots = \omega_6 = 1$. Therefore $\mathrm{Tr}(\mathbf{M}'')$ is actually the residual entropy of square ice [13, 14], and the result is obtained immediately from Eq. (2)

$$\lim_{m\to\infty} \frac{1}{m} \ln\left[\mathrm{Tr}\left(\mathbf{M}''\right)\right] = \frac{3}{2}\ln\left(\frac{4}{3}\right) = 0.431523 \ . \qquad \text{(A1)}$$